# ENTROPY OF STARS, BLACK HOLES AND DARK ENERGY

## C. SIVARAM

### Indian Institute of Astrophysics, BANGALORE

When a star collapses to form a black hole, its entropy increases considerably, for a solar mass black hole, there is a factor of $10^{19}$ increase in entropy. This corresponds to a tremendous loss of information as only the total mass, angular momentum or electric charge (if any) of the matter going inside the horizon can be measured by an outside observer. Information about all other characteristics of the matter becomes irrelevant.

For a star of mass M, contracting to a radius R, the temperature to which the matter gets heated up is given by the virial theorem as –

$$T = \frac{GMm_P}{k_B R} \quad \ldots (1)$$

(Where, $m_p$ is the proton mass, G the gravitational constant and $k_B$ the Boltzmann constant). We assume that it is constituted mainly of hydrogen, otherwise only a 'molecular factor 'of around (A/Z), enters.

Equation (1) implies

$$(RT)_{Star} = \frac{GMm_P}{k_B} \quad \ldots (2)$$

For a black hole of mass M, and Schwarzschild radius $R_S$, the Hawking temperature associated with the horizon, $T_H$ is given as:

$$R_S T_H = \frac{\hbar c}{k_B} \quad \ldots (3)$$

Consider Equation (2), one can write, $M = nm_P$, where N, the number of nucleons for a typical star is well known to be expressible as:

$$N \approx \left(\frac{\hbar c}{G m_P^2}\right)^{3/2} \qquad \ldots (4)$$

But equation (2) can be written as:

$$(RT)_{Star} = \left(\frac{\hbar c}{G m_P^2}\right)^{3/2} \left(\frac{G m_P^2}{\hbar c}\right) \left(\frac{\hbar c}{k_B}\right) = \left(\frac{\hbar c}{G m_P^2}\right)^{1/2} \left(\frac{\hbar c}{k_B}\right) \qquad \ldots (5)$$

Comparing (3) and (4), we see that there is a factor of $\left(\frac{\hbar c}{G m_P^2}\right)^{1/2} \approx 10^{19}$ in the ratio of

$$(RT)_{Star} / R_S T_H$$

This is precisely the factor by which the entropy increases. For a collapsing star, in equation (2), if we substitute $R \approx \frac{GM}{c^2}$, we get, $T \approx \frac{m_P c^2}{k_B}$, which is precisely $\left(\frac{\hbar c}{G m_P^2}\right)^{1/2} \approx 10^{19}$, times the Hawking temperature.

As the total energy (E) is the same, we have an understanding of the increase in entropy as the Hawking temperature for the horizon measured by a distant observer is lower by a factor of $\left(\frac{\hbar c}{G m_P^2}\right)$ (as compared to that measured by a local observer at the horizon), and as entropy is ~ E/T, we have a $10^{19}$ increase in entropy. A classical black hole would have $T_H = 0$, (the Hawking temperature is purely due to quantum effects) so there would complete information blackout, so the entropy increase would be indefinitely large.
(See refs (1), and (2) for a full discussion of these subtle aspects)

We considered above the case of a black hole formed by the collapse of a star composed of baryonic matter. We could also take the case of gravitating dark matter (DM) particles collapsing to form a black hole. If one typical DM particle has a mass of $m_D$, then the typical upper limit of the gravitating mass $M_{DM}$ above which it collapses to a black hole is:

$$M_{DM} \sim \frac{m_{pl}^3}{m_D^2} \qquad \ldots (6)$$

Where $m_{Pl}$ is the Planck mass, given by:

$$m_{pl} \sim \left(\hbar c/G\right)^{1/2} = 2 \times 10^{-5} \, gm \qquad \ldots (7)$$

For a baryonic star (made up of our ordinary matter), $m_p$ would take the place of $m_D$, so that the typical stellar mass before it collapses to a B.H is

$$M \sim \frac{m_{pl}^3}{m_P^2} \sim \left(\frac{\hbar c}{Gm_P^2}\right)^{3/2} m_P \sim 2M_\odot,$$

Which is indeed what we has earlier as given by equations (4) and (5)!

So a configuration of DM, before it collapses into a black hole (BH) would have entropy of (in units of $k_B$)

$$S_{DM} \approx \left(\frac{m_{pl}}{m_D}\right)^3 \qquad \ldots (8)$$

And after it collapses to a B.H, would have an entropy $S_{BH(D)}$ of:

$$S_{BH(D)} \approx \frac{GM^2}{\hbar c} \approx \left(\frac{M}{m_{pl}}\right)^2 \approx \left(\frac{m_{pl}}{m_D}\right)^4 \qquad \ldots (9)$$

(Using equation (6))

Thus when a DM configuration, made up of DM particles of mass $m_D$, collapses to form a black hole, its entropy increases (as seen from equations (8) and (9)) by a factor (F) of

$$F \sim \left(\frac{m_{pl}}{m_D}\right) \qquad \ldots (10)$$

Independent of total mass M!

For a star made up of baryons of mass $m_P$, equation (10) implies:

$$F \sim \left(\frac{m_{pl}}{m_D}\right) \sim 10^{19}, \text{ Precisely that obtained earlier in equations (4) and (5)}$$

(That is $\left(\frac{m_{pl}}{m_P}\right) = \left(\frac{\hbar c}{Gm_P^2}\right)^{1/2} \sim 10^{19}$ )

The smaller the mass of the dark matter particle, the larger will be the increase in entropy as the gravitating object made up of these particles collapses to form a black hole, the factor being ($m_{Pl}/m_{DM}$)

What about dark energy (DE)

Current cosmological observations imply a universe dominated by dark energy with its associated negative pressure. There are even indications that this may be increasing with epoch, which could result ultimately in a Big Rip and disruption of all bound structures, Dark Energy (DE) is characterized by negative pressure $P = -wc^2\rho$, W=1, corresponding to the cosmological constant introduced by Einstein, DE leads to negative gravity as acceleration can be written as:

$$a = -\frac{GM}{R^2} = -G\left(\rho + \frac{3P}{c^2}\right)\frac{4\pi R^3}{3R^2} = -G\left(\rho + \frac{3P}{c^2}\right)\frac{4\pi}{3}R \qquad \ldots (11)$$

(In GR, pressure also contributes to gravity. So $P = -\rho c^2$, implies a, positive, i.e. not deceleration as in the case of attractive gravity, but repulsion. For the present, we will assume a constant dark energy density, i.e. essentially a cosmological constant Λ, with

$$P_\Lambda = \frac{\Lambda c^2}{8\pi G} \qquad \ldots (12)$$

At present, DE constitutes at least 0.7 of the material in the universe as implied by supernovae (SN) and WMAP observations.

The space-time, which describes a universe almost completely dominated by DE, is the de-sitter universe, with constant curvature Λ, ($G_{\mu\nu} = \Lambda g_{\mu\nu}$)

The de sitter space (with positive Λ), has a horizon given by

$$r = \sqrt{3/\Lambda}$$

It is a space of constant curvature, and a remarkable property of such a space is that it is 'hot', i.e., has a horizon temperature of

$$T = \frac{\hbar c}{k_B} \sqrt{\Lambda} \qquad \ldots (13)$$

(In GR, curvature is measured by relative acceleration between neighbouring particles moving along geodesics and by the Unruh – Davies effect an accelerated observer (acceleration a) would see himself surrounded by a thermal bath of temperature $T \sim \frac{\hbar a}{k_B c}$. As a consequence of the equivalence principle this is just the Hawking temperature, for a black hole the surface gravity corresponding to acceleration!)

As the universe becomes increasingly dominated by DE, we can approximate the overall space-time as that of de-sitter.

The volume of the de sitter space is $\sim 2\Pi^2 \Lambda^{-3/2}$, so that using equations (12) and (13), we can get the total entropy of the D.E as:

$$S_{univ} \approx S_{DE} \approx \frac{c^3 k_B}{\hbar G \Lambda} \qquad \ldots (14)$$

This can be written as:SS

$$S_{DE} = \frac{k_B}{\left(\hbar G / c^3\right)\Lambda} = \frac{k_B}{L_{pl}^2 \Lambda} \qquad \ldots (15)$$

This shows that the entropy associated with the (DE dominated) space of curvature $\Lambda$, is $\sim 1/\Lambda$. For constant $\Lambda$, the associated entropy remains invariant.

(For $\Lambda \sim 10^{-56} cm^{-2}$, as implied by current observations, $L_{pl}^2 \sim 10^{66}$ cm $^{-2}$) we have

$$S_{DE} \cong 10^{122} K_B! \qquad \ldots (16)$$

This is the asymptotic entropy of the universe as it becomes completely DE dominated. What would be the black hole mass corresponding to the entropy given by equation (16)? We have

$$\frac{k_B}{L_{pl}^2 \Lambda} = \left(\frac{M}{m_{pl}}\right)^2 k_B$$

$$M^2 \approx \frac{m_{pl}^2}{L_{pl}^2 \Lambda},$$

$$M \sim 10^{60} M_{pl} \sim 10^{22} M_\Theta$$

This corresponds roughly to the total mass of all black holes in the universe.

In a recent work (ref.5) (see also earlier section) we have shown that the total entropy released from the (dark energy induced) ripping of all the black holes matches with the entropy of the asymptotic D.E dominated S.T suggesting links between the two major conundrums, i.e., D.E and B.H entropy. There are other remarkable aspects to equation (15). For a black hole of mass M, we can replace $\Lambda$, (in equation 15) by the Schwarzschild curvature of the horizon, $u \sim \left(c^4 / G^2 M^2\right)$; we then get just the black hole entropy! This implies this is a universal type of relation. Again if the universe began with a high curvature, $\Lambda \sim 1/L_{pl}^2$, equation (15) implies that it would have had an initial entropy of just $k_B$ ! This is a minimal value for entropy, just as h is the minimal action. As the curvature of the universe decreased as it expanded, the entropy keeps increasing.

So if $\Lambda$ scales as $t^{-2}$, in the initial stages, the entropy would have grown with the expansion time as $t^2$, so that the entire observed entropy could have been generalized in a time $\sim 10^{44}$ Planck time! (< 1 second)

All this is entirely consistent with the second law of thermodynamics, which requires increasing entropy and a growing cosmological time arrow. A varying DE can also be accommodated in the above discussion.

A recent model for growing DE is realized for a scalar field with potential where $m_d \sim 10^{-32}$ eV. (ref.6).

This is precisely what equation (9) above combined with equation (15) gives!